\documentclass[preprint,tightenlines,showpacs,amsmath,amssymb]{revtex4}
\usepackage{graphicx}% Include figure files
\usepackage{dcolumn}% Align table columns on decimal point
\usepackage{bm}% bold math
\usepackage{rotating}
\newcommand{\psip}{\psi^\prime}

\newcommand{\psp}{\psi^\prime}
\newcommand{\jpsi}{J/\psi}

\newcommand{\chicz}{\chi_{c0}}
\newcommand{\chico}{\chi_{c1}}
\newcommand{\chict}{\chi_{c2}}
\newcommand{\EE}{e^+e^-}
\newcommand{\MM}{\mu^+\mu^-}

\newcommand{\pp}{\pi^+\pi^-}

\newcommand{\ppb}{p\overline{p}}

\newcommand{\pspto}{\psi^\prime \rightarrow }

\newcommand{\chictto}{\chi_{c2} \rightarrow }
\newcommand{\bfg}{\begin{figure}}
\newcommand{\efg}{\end{figure}}
\newcommand{\bitm}{\begin{itemize}}
\newcommand{\eitm}{\end{itemize}}
\newcommand{\bnum}{\begin{enumerate}}
\newcommand{\enum}{\end{enumerate}}
\newcommand{\btbl}{\begin{table}}
\newcommand{\etbl}{\end{table}}
\newcommand{\btbu}{\begin{tabular}}
\newcommand{\etbu}{\end{tabular}}

\newcommand{\GG}{\gamma\gamma}
\newcommand{\kk}{K^+K^-}

\newcommand{\LL}{\ell^+\ell^-}
\newcommand{\beq}{\begin{equation}}
\newcommand{\edq}{\end{equation}}
\newcommand{\g}{\gamma}

\begin{document}
\normalsize
\parskip=5pt plus 1pt minus 1pt

%\preprint{} \preprint{%\vbox{
                      %  \hbox{Intended for {\it Phys. Rev. D}}
                      %  \hbox{Authors: Z. Q. Liu, C. P. Shen, C. Z. Yuan, F. A. Harris and G. Li }
                      %  \hbox{Committee: S.~S. Fang (chair), J.~M. Bian, D.~Cronin-Hennessy} }
                      %}

\title{\quad\\[1.0cm] \boldmath Higher-order multipole amplitude measurement in $\pspto\g\chict$}

%\author{Author list}
%\begin{small}
%\begin{center}
\author{
{\small
M.~Ablikim$^{1}$, M.~N.~Achasov$^{5}$, D.~Alberto$^{40}$, F.~F.~An$^{1}$, Q.~An$^{38}$, Z.~H.~An$^{1}$, J.~Z.~Bai$^{1}$, R.~Baldini$^{19}$, Y.~Ban$^{25}$, J.~Becker$^{2}$
% ...
\\
\vspace{0.2cm}
(BESIII Collaboration)\\
\vspace{0.2cm} {\it
$^{1}$ Institute of High Energy Physics, Beijing 100049, P. R. China\\
$^{2}$ Bochum Ruhr-University, 44780 Bochum, Germany\\
% ...
$^{5}$ G.I. Budker Institute of Nuclear Physics SB RAS (BINP), Novosibirsk 630090, Russia\\
% ...
$^{19}$ Laboratori Nazionali di Frascati - INFN, 00044 Frascati, Italy\\
% ...
$^{25}$ Peking University, Beijing 100871, P. R. China\\
% ...
$^{38}$ University of Science and Technology of China, Hefei 230026, P. R. China\\
% ...
$^{40}$ University of Turin and INFN, Turin, Italy\\
% ...
}}
%\end{center}
\vspace{0.4cm}
\vspace{8cm}
}
%\end{small}

%\date{\today}

\begin{abstract}

Using $106\times10^6$ $\psp$ events collected with the BESIII detector
at the BEPCII storage ring, the higher-order multipole amplitudes in
the radiative transition $\pspto\g\chictto\g\pp/\g\kk$ are measured. A
fit to the $\chict$ production and decay angular distributions yields
$M2=0.046\pm0.010\pm0.013$ and $E3=0.015\pm0.008\pm0.018$, where the
first errors are statistical and the second systematic. Here
$M2$ denotes the
normalized magnetic quadrupole amplitude and $E3$ the normalized
electric octupole amplitude. This measurement shows evidence for the
existence of the $M2$ signal with $4.4\sigma$ statistical significance
and is consistent with the charm quark
having no anomalous magnetic moment.

\end{abstract}


\maketitle

\section{Introduction}

The radiative transitions $\pspto\g\chi_{cJ}$ $(J=1,2)$ provide
information about the electromagnetic interaction between charm and
anti-charm quarks in charmonia and allow investigation of many
interesting topics, including whether the charm quark has an anomalous
magnetic moment~\cite{mult-pole,prediction} or if there is S-wave and
D-wave state mixing~\cite{wave-mix}.  In general, the transition
amplitude of radiative decays of charmonium states is dominated by the
electric dipole (E1) contribution, with higher multipoles suppressed
by powers of photon energy divided by charm quark
mass~\cite{quarkmass}.  The search for contributions of higher-order
multipole amplitude is of interest as a source of information on the
charm quark's magnetic moment; the possibility of anomalous
magnetic moments of heavy quarks being larger than those of lighter
ones was raised in Ref.~\cite{wilson}. In $\pspto \gamma \chi_{c2}$,
taking the charm quark to have a mass $m_c=1.5$ GeV/$c^2$ and an
anomalous magnetic moment $\kappa$,
$M2=0.029(1+\kappa)$ is predicted~\cite{quarkmass}.

Disagreement between a pure E1 calculation and experimental
measurements~\cite{conflict} hint that higher order multipole
amplitudes may exist.  These would be reflected in the angular
distributions of both the radiative photon and the final state
particles~\cite{angle,angle-corr}. Thus, careful investigation of the
angular distributions is important.

Several experiments, including the Crystal Ball experiment in
$\pspto\g\chi_{c1,c2}\to\GG\jpsi\to\GG\LL (\ell=e~ \hbox{or}~
\mu)$~\cite{crystal}, the E-835 experiment in $\ppb\to\chi_{c1,c2}
\to\g\jpsi\to\g\EE$~\cite{E835}, the E-760 experiment in
$\ppb\to\chictto\g\jpsi \to\g\EE$~\cite{E760}, and the BESII
experiment in $\pspto\g\chictto\g\pp/\g\kk$~\cite{bes2} have
searched for higher-order multipole amplitudes. Due to their limited
statistics, they were unable to provide evidence for the existence of
higher-order multipoles. More recently, the CLEO experiment reported
measurements of higher-order multipole amplitudes in
$\pspto\g\chi_{c1,c2}\to\GG\jpsi\to\GG\LL$~\cite{cleo}, where
significant $M2$ contributions were found in the $\pspto\g\chico$ and
$\chi_{c1,c2}\to\g\jpsi$ transitions. Tables~\ref{result1} and
\ref{result2} summarize the experimental measurements on searches for
higher-order multipole amplitudes.

%%%%%%%%%%%%%%%
\begin{table}[htb]
\caption{Current experimental measurements of the normalized M2
contributions in the decays $\chi_{c1}\to\gamma
J/\psi$  and $\psi^\prime\to\gamma\chi_{c1}$.}
\label{result1}
\begin{center}
  \begin{tabular}{lccc}
  \hline\hline
  Experiment & $\chi_{c1}\to\gamma
J/\psi$  & $\psi^\prime\to\gamma\chi_{c1}$ & Signal Events \\
  \hline
  Crystal Ball{~\cite{crystal}} & $-0.002_{-0.020}^{+0.008}$ & $0.077_{-0.045}^{+0.050}$ & 921\\
  E-835{~\cite{E835}} & $0.002\pm0.032\pm0.004$ & - & 2090\\
  CLEO-c{~\cite{cleo}} & $-0.0626\pm0.0063\pm0.0024$ & $0.0276\pm0.0073\pm0.0023$ & 39363\\
  \hline
  \end{tabular}
\end{center}
\end{table}

% This is Table II
\begin{table}[htb]
\caption{Current experimental measurements of the normalized
M2 contributions in the decays $\chi_{c2}\to\gamma
J/\psi$  and $\psi^\prime\to\gamma\chi_{c2}$.}
\label{result2}
\begin{center}
  \begin{tabular}{lccc}
  \hline \hline
  Experiment & $\chi_{c2}\to\gamma
J/\psi$  & $\psi^\prime\to\gamma\chi_{c2}$ & Signal Events\\
  \hline
  Crystal Ball {\cite{crystal}} & $-0.333_{-0.292}^{+0.116}$ & $0.132_{-0.075}^{+0.098}$ & 441\\
  E-760 {~\cite{E760}} & $-0.14\pm0.06$ & - & 1904\\
  E-835 {~\cite{E835}} & $-0.093_{-0.041}^{+0.039}\pm0.006$ & - & 5908\\
  BESII {~\cite{bes2}} & - & $-0.051_{-0.036}^{+0.054}$ & 731\\
  CLEO-c{~\cite{cleo}} & $-0.079\pm0.019\pm0.003$ & $0.002\pm0.014\pm0.004$ & 19755\\
  \hline
  \end{tabular}
\end{center}
\end{table}
%%%%%%%%%%%%%%%

In this article, $(1.06\pm0.04)\times10^8$ $\psip$
events~\cite{psipnumber} accumulated in the BESIII experiment are used
in the selection of $\pspto\g\chict, \chict\to\pp/\kk$ events, which
allow the determination of the higher-order multipole amplitudes in
the $\pspto\g\chict$ transition.

\section{The BESIII experiment and data set}

This analysis is based on a $156.4$ pb$^{-1}$ of $\psip$ data
corresponding to $(1.06\pm0.04)\times10^8$ $\psip$
events~\cite{psipnumber} collected with the BESIII
detector~\cite{bes3} operating at the BEPCII Collider~\cite{bepc}.  In
addition, an off-resonance sample of $42.6$ pb$^{-1}$ taken at
$\sqrt{s}=3.65$ GeV is used for the study of continuum backgrounds.

BESIII/BEPCII~\cite{bes3} is a major upgrade of the BESII experiment
at the BEPC accelerator~\cite{bepc} for studies of hadron spectroscopy
and $\tau$-charm physics~\cite{bes3yellow}. The design peak luminosity
of the double-ring $\EE$ collider, BEPCII, is $10^{33}$
cm$^{-2}$s$^{-1}$ at beam currents of 0.93 A. The BESIII detector
with a geometrical acceptance of 93\% of 4$\pi$, consists of the
following main components: 1) a small-celled, helium-based main draft
chamber (MDC) with 43 layers. The average single wire resolution is
135 $\mu$m, and the momentum resolution for 1 GeV/c charged particles
in a 1 T magnetic field is 0.5\%; 2) an electromagnetic calorimeter
(EMC) made of 6240 CsI (Tl) crystals arranged in a cylindrical shape
(barrel) plus two end-caps. For 1.0 GeV photons, the energy resolution
is 2.5\% in the barrel and 5\% in the endcaps, and the position
resolution is 6 mm in the barrel and 9 mm in the end-caps; 3) a
Time-Of-Flight system (TOF) for particle identification (PID) composed
of a barrel part made of two layers with 88 pieces of 5 cm thick, 2.4
m long plastic scintillators in each layer, and two end-caps with 48
fan-shaped, 5 cm thick, plastic scintillators in each end-cap. The time
resolution is 80 ps in the barrel, and 110 ps in the end-caps,
corresponding to a $K/\pi$ separation by more than 2$\sigma$ for
momenta below about 1 GeV/c; 4) a muon chamber system (MUC) made of
1000 m$^2$ of Resistive Plate Chambers (RPC) arranged in 9 layers in
the barrel and 8 layers in the end-caps and incorporated in the return
iron yoke of the superconducting magnet.  The position resolution is
about 2 cm.

The optimization of the event selection and the estimation of physics
backgrounds are performed through Monte Carlo (MC) simulations. The
GEANT4-based simulation software BOOST~\cite{boost} includes the
geometric and material description of the BESIII detectors and the
detector response and digitization models, as well as the tracking of
the detector running conditions and performance. The production of the
$\psip$ resonance is simulated by the MC event generator
KKMC~\cite{kkmc}, while the decays are generated by
EVTGEN~\cite{evtgen} for known decay modes with branching ratios being
set to PDG~\cite{pdg} world average values, and by
LUNDCHARM~\cite{lundcharm} for the remaining unknown decays. The
analysis is performed in the framework of the BESIII offline software
system~\cite{boss} which takes care of the detector calibration, event
reconstruction and data storage.

MC samples of $\psip\to\gamma\chi_{c0,c2}\to\gamma\pp/\gamma\kk$ are
generated according to phase space to determine the normalization
factors in the partial wave analysis~\cite{bes2,shix-thesis}, and MC
samples of $\psip\to(\gamma)\EE/(\gamma)\MM$ and $\psip\to X\jpsi$
($X=\pi^0\pi^0,\eta$) with $\jpsi\to(\gamma)\MM$ are generated for
background studies.

\section{Data analysis}
Charged tracks are reconstructed in the MDC, and the number of charged
tracks is required to be two and have no net charge. For each track, the
polar angle must satisfy $|\cos\theta|<0.93$, and the point of closest
approach must be within $\pm10$ cm of the interaction point in the
beam direction and within $\pm1$ cm of the beam line in the plane
perpendicular to the beam.  The TOF (both end-cap and Barrel) and
$dE/dx$ measurements for each charged track are used to calculate
$\chi_{PID}^2(i)$ values and the corresponding confidence levels
$Prob_{PID}(i)$ for the hypotheses that a track is a pion, kaon, or
proton, where $i~(i=\pi/K/p)$ is the particle type. For pion
candidates, $Prob_{PID}(\pi)>0.001$ is required, while for kaon
candidates, $Prob_{PID}(K)>Prob_{PID}(\pi)$ and $Prob_{PID}(K)>0.001$
are required.

Electromagnetic showers are reconstructed by clustering EMC crystal
energies. The energy deposited in nearby TOF counters is included to
improve the reconstruction efficiency and energy resolution. Showers
identified as photon candidates must satisfy fiducial and
shower-quality requirements.  The minimum energy is 25 MeV for barrel
showers ($|\cos\theta|<0.8$) and 50 MeV for end-cap showers
($0.86<|\cos\theta|<0.92$). The showers in the angular range between
the barrel and end-cap are poorly reconstructed and excluded from the
analysis.  To eliminate showers from charged particles, a photon must
be separated by at least $20$ degrees from any charged track.
EMC cluster timing requirements are used to suppress electronic noise
and energy deposits unrelated to the event.  The number of good photon
candidates is required to be larger than or equal to one in each event,
and the photon with the highest energy is regarded as the radiative
photon from $\psip\to\gamma\chict$.

In order to separate pions and kaons more effectively and to
distinguish $\chicz$ and $\chict$ more clearly, a four-constraint
kinematic fit (4-C fit) is performed with the two charged tracks and
the radiative photon candidate under the hypotheses that the two tracks
are either $\pp$ or $\kk$, and the kinematic chi-squares, $\chi^2_\pi$
and $\chi^2_K$, are determined. If $\chi^2_\pi<\chi^2_K$ and
$\chi^2_\pi<60$, the event is categorized as $\gamma\pp$; otherwise,
if $\chi^2_K<\chi^2_\pi$ and $\chi^2_K<60$, the event is categorized
as $\gamma\kk$. For the selected $\gamma\pp$ and $\gamma\kk$ candidate
events, at least one of the charged tracks is required to be identified as
a $\pi$ for $\gamma\pp$ or a $K$ for $\gamma\kk$.

To remove $\EE\to(\gamma)\EE$ and $\psip\to(\gamma)\EE$ backgrounds,
the deposited energy of each track in the EMC is required to be less
than $1.4$ GeV, and the observed ionization is also required to be
within $3\sigma$ of the expected value for each track. Furthermore, in
$\gamma\pp$, observed ionization of charged tracks is required to be
within $2\sigma$ of the expected value when the polar angle of the
charged track is within the EMC insensitive region
(i.e. $0.81<|\cos\theta|<0.86$). These requirements remove almost all
events with two electron tracks but still keep the efficiencies for
the signal channels very high: 96\% for $\gamma \pp$ and 97\% for
$\gamma \kk$.

In $\gamma\pp$, there are $\EE\to(\gamma)\MM$ and
$\psip\to(\gamma)\MM$ backgrounds due to $\pi/\mu$
misidentification. In order to remove the backgrounds with $\MM$, the
deposited energy in the EMC of at least one of the charged tracks is
required to be larger than $0.34$ GeV. Over $99\%$ of the
$\EE\to(\gamma)\MM$ and $\psip\to(\gamma)\MM$ backgrounds are removed
after applying this requirement. Since $\mu/K$ misidentification
is quite small, it is not necessary to apply this requirement in the
$\gamma\kk$ decay.

\begin{figure}
\begin{center}
\includegraphics[height=5cm]{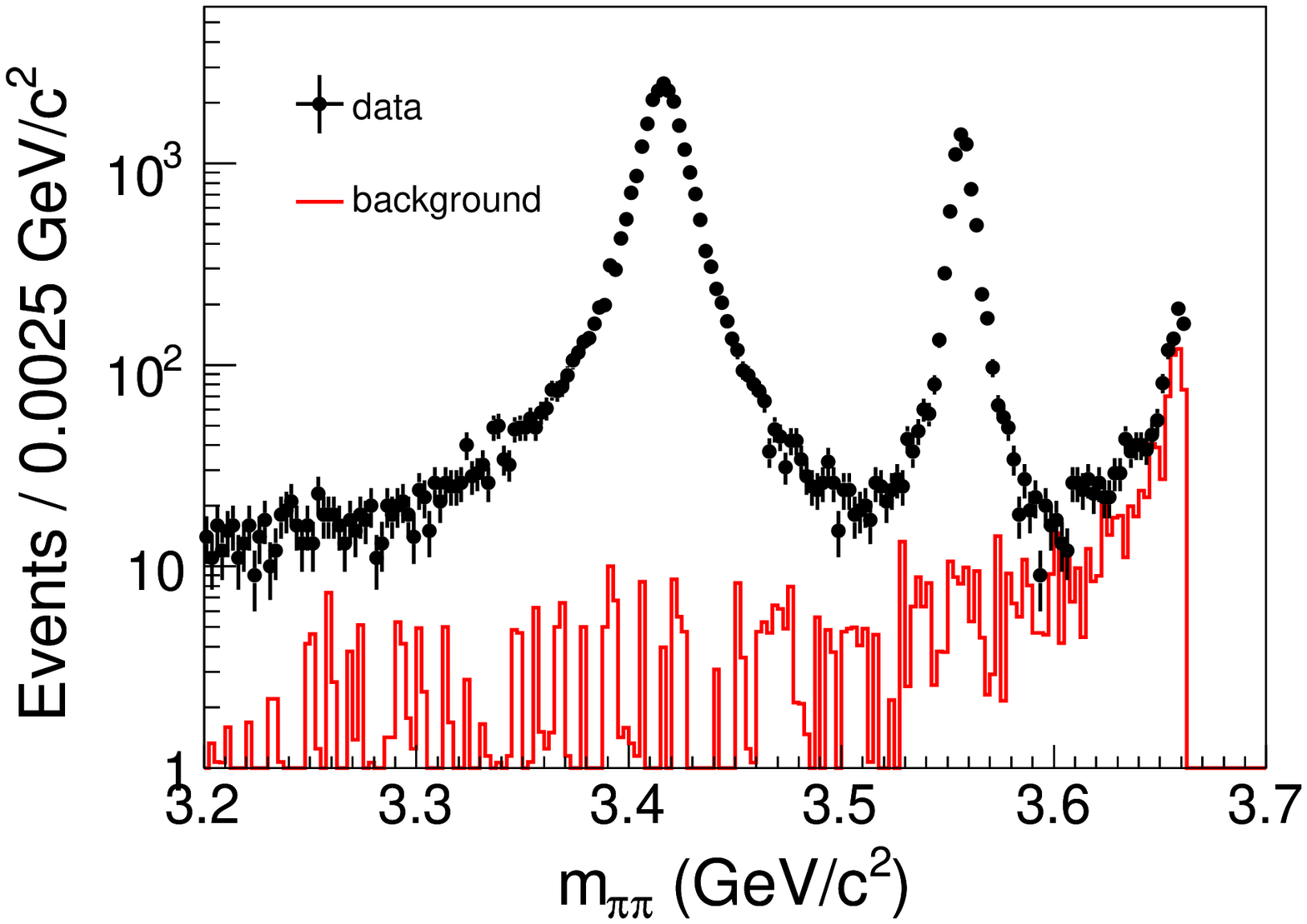}
\includegraphics[height=5cm]{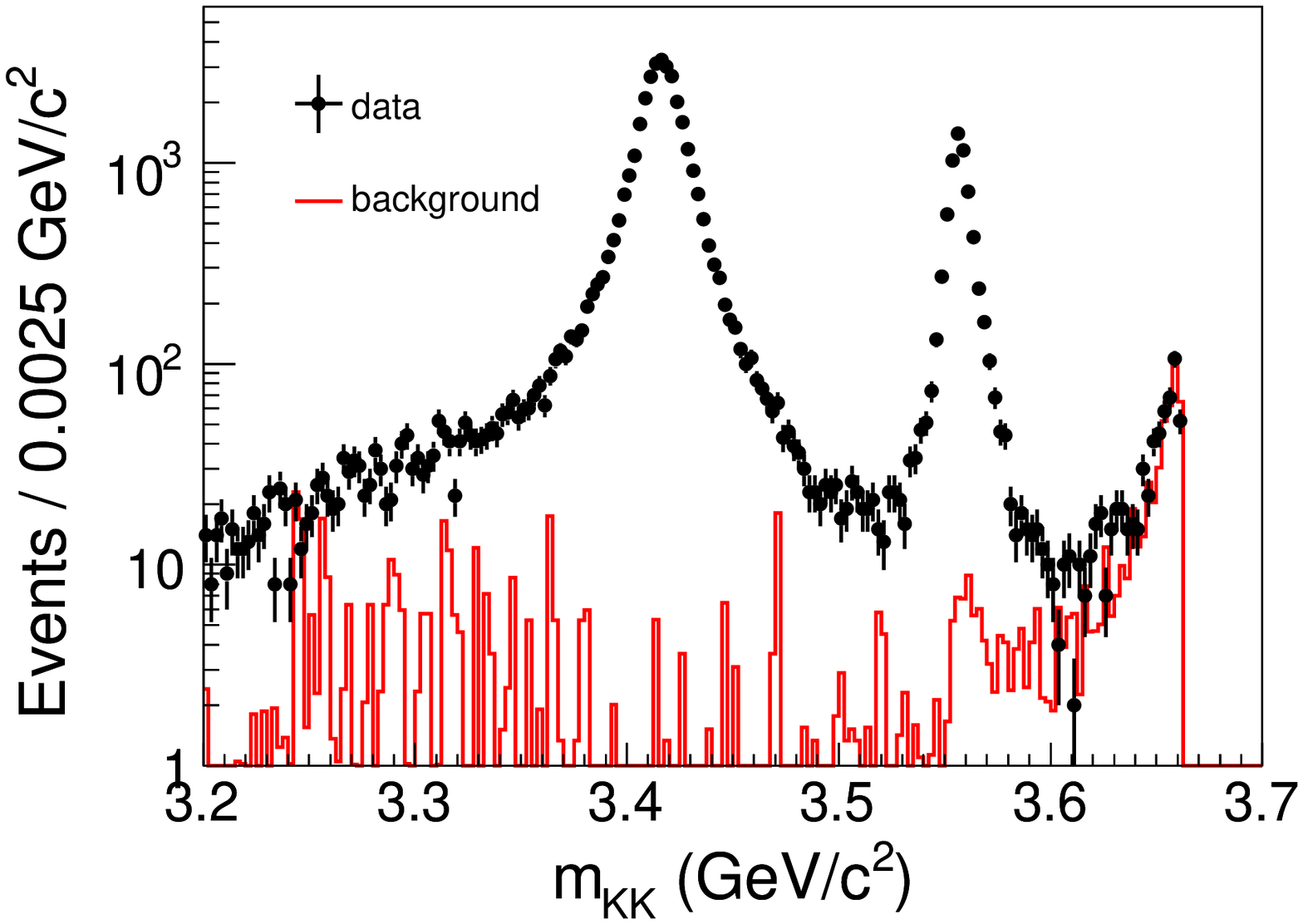}
\caption{The invariant mass distributions of $\pp$ (left) and $\kk$
(right) for the selected $\gamma \pp$ and $\gamma \kk$ events from
$\psip$ data. Dots with error bars are data while blank
histograms are the sum of MC simulated backgrounds and normalized
continuum background (estimated from the data sample taken at $\sqrt{s}=3.65$
GeV).} \label{mass}
\end{center}
\end{figure}

After performing all the above selection criteria, clean
$\psip\to\gamma\chi_{c0,c2}\to\gamma\pp/\gamma\kk$ data samples are
obtained. The $\pp$ and $\kk$ invariant mass distributions are shown
in Fig.~\ref{mass}.  Clear $\chicz$ and $\chict$ signals are observed,
and the background levels within the $\chi_{c2}$ signal region between
3.53 and 3.59 GeV/$c^2$ are 2.6\% 
[0.7\% $\psip\to(\gamma)\MM$,
0.8\% normalized continuum, 0.3\% cross contamination from
$\chi_{c2}\to\kk$, 0.7\% $\chi_{c0}$ tail, and 0.1\%
$\psip\to\pp/\pp\pi^0$ events]
for $\gamma \pp$ and 2.1\% 
[0.7\% cross contamination from $\chi_{c2}\to\pp$, 1.1\%
$\chi_{c0}$ tail, and 0.3\% $\psip\to\kk$ events]
for $\gamma \kk$.
The highest mass peak corresponds to $\psip$ decays to
two charged tracks that are kinematically fitted with an
unassociated low energy photon.
Requiring the
invariant mass of the two charged tracks to be between $3.53$ and
$3.59$ GeV/c$^2$ to select $\chict$, $7154$ $\gamma\pp$ events
and $6657$ $\gamma\kk$ events are obtained.

%%%%%%%%%% invariant mass %%%%%%%%%
%%%%%%%%%%%%%%%%%%%%%%%%%%%%%%%%%%%

\section{Fit to the angular distributions}

The formulae for the helicity amplitudes in $\psip\to\gamma\chict\to\gamma PP$
($P=\pi/K$), which include higher-order multipole amplitudes,
are widely discussed in Refs.~\cite{angle,angle-corr,bes2}:
%%%%%% helicity formula %%%%%%%%
\begin{eqnarray}\label{eq-angle}
\notag W(\theta_{\gamma},\theta_{M},\phi_{M}) &=&
3\sin^{2}\theta_{\gamma}\sin^{2}(2\theta_{M})x^{2}\\\notag &+&
\frac{3}{2}(1+\cos^{2}\theta_{\gamma})\sin^{4}\theta_{M}y^{2}\\\notag
&-&
\frac{3}{\sqrt{2}}\sin(2\theta_{\gamma})\sin(2\theta_{M})\sin^{2}\theta_{M}\cos\phi_{M}xy\\\notag
&+&
\sqrt{3}\sin(2\theta_{\gamma})\sin(2\theta_{M})(3\cos^{2}\theta_{M}-1)\cos\phi_{M}x\\\notag
&+&
\sqrt{6}\sin^{2}\theta_{\gamma}\sin^{2}\theta_{M}(3\cos^{2}\theta_{M}-1)\cos2\phi_{M}y\\
&+& (1+\cos^{2}\theta_{\gamma})(3\cos^{2}\theta_{M}-1)^2\\\notag
\end{eqnarray}
%%%%%%%%%%%%%%%%%%%%%%%%%%%%%%%
where $x=A_1/A_0,~y=A_2/A_0$, and $A_{0,1,2}$ are the $\chict$
helicity $0,1,2$ amplitudes, respectively. $\theta_\gamma$ is the
polar angle of the radiative photon where the electron beam is defined
as the $z$ axis in the $e^+e^-$ center-of-mass frame, and $\theta_M$
and $\phi_M$ are the polar and azimuthal angles of the $\pi/K$ in the
$\chi_{c2}$ rest frame, where the polar axis is defined with respect
to the radiative photon direction and $\phi_M=0$ is defined by the
electron beam direction.

An unbinned maximum likelihood fit to the joint production and decay
angular distribution is performed to determine $x$ and $y$ values. We
define six factors~\cite{shix-thesis}:
\begin{eqnarray}
a_1 &=& 3\sin^{2}\theta_{\gamma}\sin^{2}(2\theta_{M}),\\
a_2 &=& \frac{3}{2}(1+\cos^{2}\theta_{\gamma})\sin^{4}\theta_{M},\\
a_3 &=& -\frac{3}{\sqrt{2}}\sin(2\theta_{\gamma})\sin(2\theta_{M})\sin^{2}\theta_{M}\cos\phi_{M},\\
a_4 &=& \sqrt{3}\sin(2\theta_{\gamma})\sin(2\theta_{M})(3\cos^{2}\theta_{M}-1)\cos\phi_{M},\\
a_5 &=& \sqrt{6}\sin^{2}\theta_{\gamma}\sin^{2}\theta_{M}(3\cos^{2}\theta_{M}-1)\cos2\phi_{M},\\
a_6 &=& (1+\cos^{2}\theta_{\gamma})(3\cos^{2}\theta_{M}-1)^2.\\\notag
\end{eqnarray}

The mean values of $a_1,a_2,a_3,a_4,a_5,a_6$ are determined with
$\psip \to \gamma\chict, \chict \to PP$ MC events, where phase space
is used for the simulation of all the angular distributions:
\begin{equation}
\bar{a_n}=\frac{\Sigma^N_{i=1}a_n(i)}{N},~~~~~~n=1,...,6
\end{equation}
Here N is the number of events after all selections from phase space MC samples.
We integrate firstly the parts independent of the parameters in the angular distribution
to make the fit faster.
Since $\bar{a_n}$ is calculated with phase space MC events, thus it accounts for the
detector acceptance effects naturally.

Then, the constructed probability-density function (pdf) is
written as:
\begin{equation}
f(x,y)=\frac{W(\theta_\gamma,\theta_M,\phi_M)}{\bar{a_1}x^2+\bar{a_2}y^2+
\bar{a_3}xy+\bar{a_4}x+\bar{a_5}y+\bar{a_6}}
\end{equation}
In practice, we use the log-likelihood
function, which is given by
$\ln\mathcal{L}=\Sigma^N_{i=1}\ln f_i(x,y)$ for convenience, where
the sum is over the events in the signal region. The dominant background
events are simulated by MC events and their normalized contributions are subtracted in
$\ln\mathcal{L}$ value, i.e. $\ln\mathcal{L}_s=\ln\mathcal{L}-\ln\mathcal{L}_b$,
where $\ln\mathcal{L}_b$ is the normalized sum of
logarithmic likelihood values from background events.

\begin{figure}
\begin{center}
\includegraphics[height=5.5in]{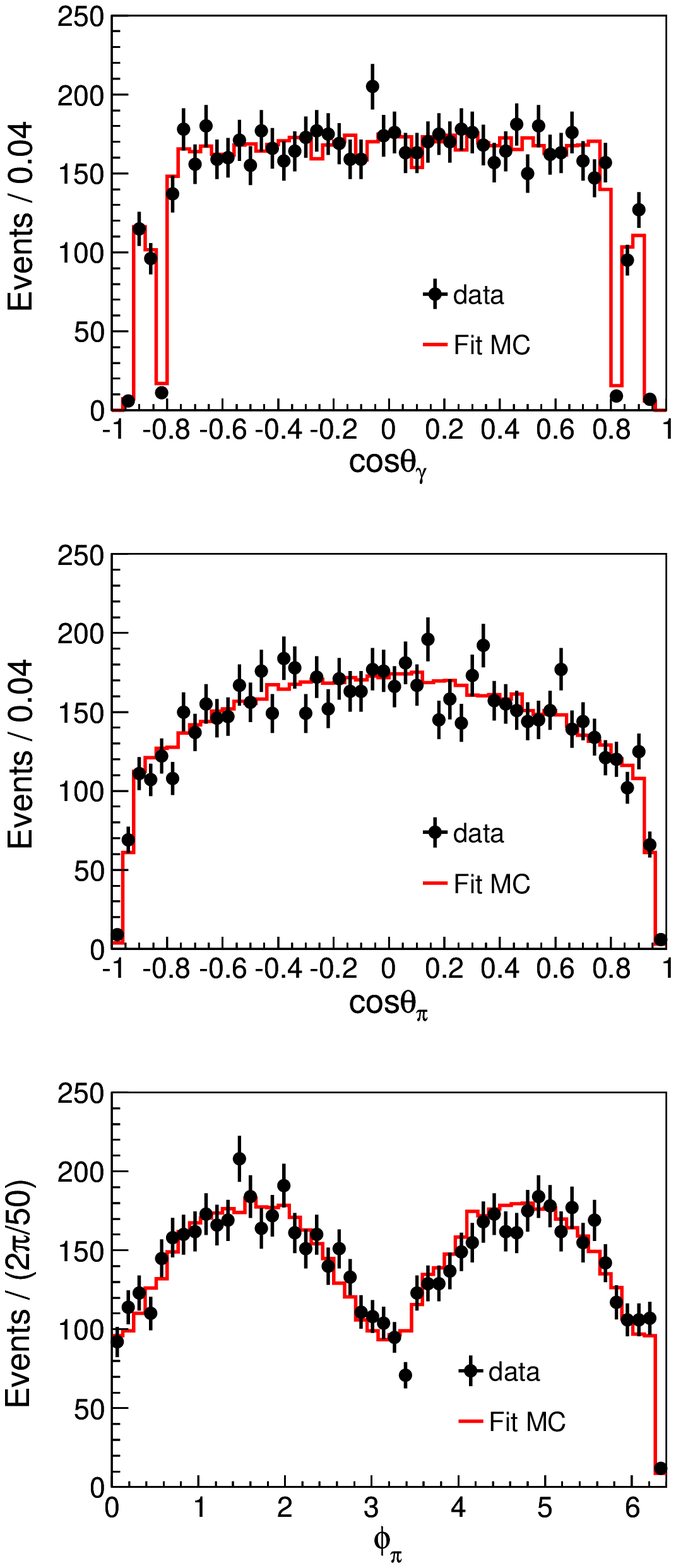}
\includegraphics[height=5.5in]{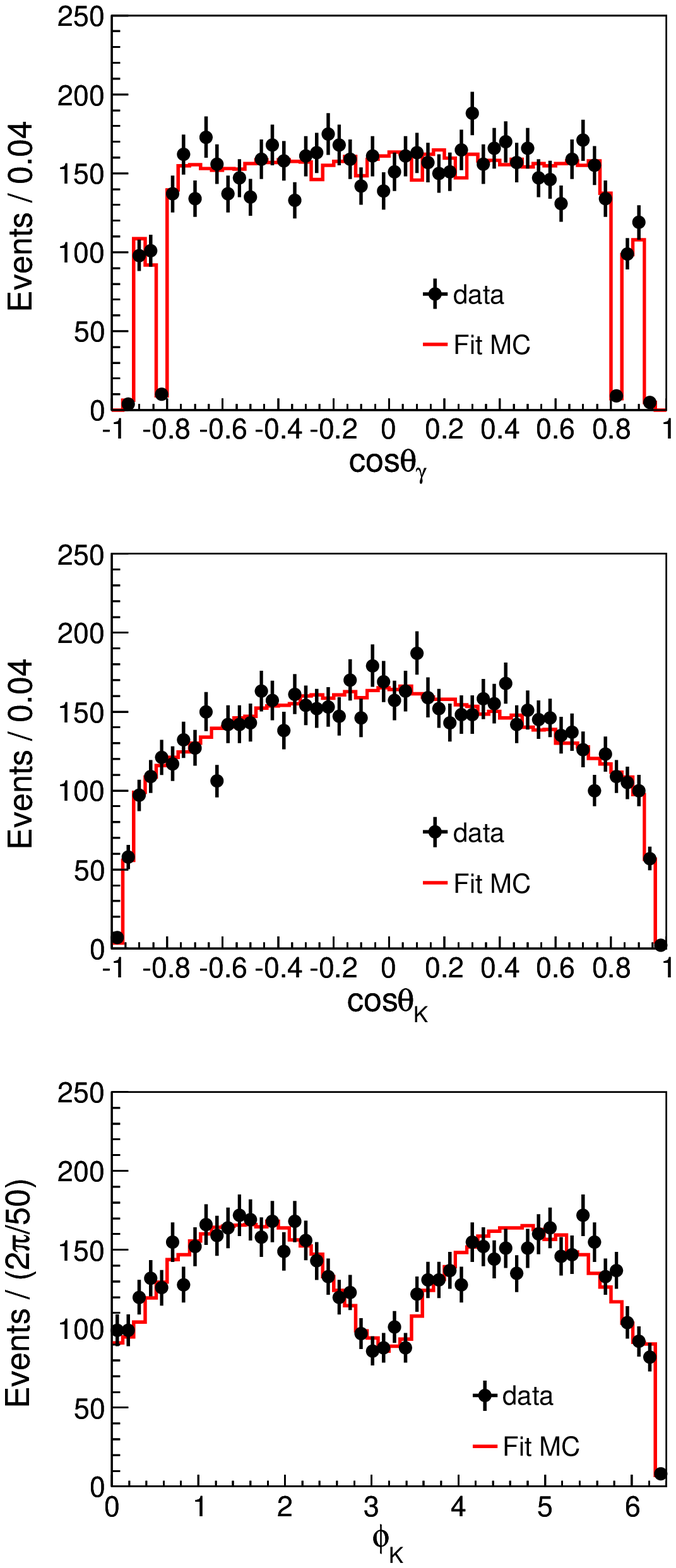}
\caption{ The angular distributions of $\cos\theta_\gamma$,
$\cos\theta_M$ and $\phi_M$ for $\chi_{c2}\to \pp$ (left) and
$\chi_{c2}\to \kk$ (right), where the dots with error bars are data
and the histograms are the fitted results.}\label{chic2-angle}
\end{center}
\end{figure}

Before fitting to the data, input and output checks have been done
using MC samples, and the checked results verify the
validity of the fitting procedure.
An unbinned maximum likelihood fit
to the $\gamma\pp$ and $\gamma\kk$ production and decay angular
distributions yield
%%%% Fit angular distribution %%%
\begin{eqnarray}\label{fit-gpp}
x_\pi=1.55^{+0.08}_{-0.07},~y_\pi=2.06^{+0.10}_{-0.09},~\rho_\pi=0.890.\\
x_K=1.55\pm0.08,~y_K=2.13^{+0.11}_{-0.10},~\rho_K=0.902.
\end{eqnarray}
%%%%%%%%%%%%%%%%%%%%%%%%%%%%%%%%%
where the errors are statistical, and $\rho_\pi,~\rho_K$ are the
correlation coefficients between $x$ and $y$ for $\gamma\pp$ and
$\gamma\kk$, respectively. A simultaneous fit to $\gamma\pp$ and
$\gamma\kk$ gives
%%% Simultaneous fit chi_c2 %%%
\begin{equation}\label{fit-simultaneous}
x=1.55\pm0.05,~y=2.10\pm0.07,~\rho=0.896
\end{equation}
%%%%%%%%%%%%%%%%%%%%%%%%%%%%%%%
where errors are statistical, and $\rho$ is the correlation
coefficient between $x$ and $y$.
The normalized M2 and E3 amplitudes are
calculated to be~\cite{mult-pole}:
\begin{equation}
M2=0.046\pm0.010,~E3=0.015\pm0.008
\end{equation}
based on the simultaneous fit result, where errors are statistical only.
Figure~\ref{chic2-angle} shows the
angular distributions of data and the fitted results both for
$\gamma\pp$ and $\gamma\kk$ events. Good agreement is observed for all
angular distributions.

The goodness of the fit is estimated using Pearson's $\chi^2$
test~\cite{chi2test}. The data sample is divided into
$8\times8\times8=512$ bins in $\cos\theta_\gamma$, $\cos\theta_M$, and
$\phi_M$, and the $\chi^2$ value is calculated as:
\begin{equation}
\chi^2=\sum_i\frac{(n_i^{DT}-n_i^{MC})^2}{n_i^{DT}}
\end{equation}
where $n_i^{DT}$ and $n_i^{MC}$ are the number of the observed events
in $i$-th bin from data and the corresponding number of normalized
events from MC using $x$ and $y$ fixed to the values determined in the
analysis. Here MC events are $20$ times more than
data events.
For bins with less than 7 events, we add the
events into the adjacent bin. The result yields
$\chi^2_\pi/n.d.f=377.5/368=1.03$ for $\gamma\pp$ and
$\chi^2_K/n.d.f=348.1/354=0.98$ for $\gamma\kk$, where $n.d.f$ is the
number of the degrees of freedom. These results show that the fits are
good.

%%%% chi_c2 angle distribution %%%%
%%%%%%%%%%%%%%%%%%

\section{Systematic Errors}

\subsection{MC simulation of detector response}
The consistency between data and MC simulation for $\chict$ events can
be tested using $\chicz$ events. The angular distribution of $\chicz$
is unambiguous, i.e., $W_0=1+\cos^2\theta_\gamma$. If we replace the
$(3\cos^2\theta_M-1)^2$ term in Eq.~1 by 1, then Eq.~1 becomes equal
to $W_0$ when both $x$ and $y$ equal zero. Therefore, if we fit the
angular distribution of $\chicz$ events with a modified Eq.~1 using
the same method as in $\chict$ decays, $x=0$ and $y=0$ are
expected. Non-zero $x$ and $y$ values from the fit reflect the
difference between data and MC and give a measure of the systematic
error due to the MC simulation of the detector response.  The fitted
results are
$x_\pi=0.049^{+0.016}_{-0.017},~y_\pi=-(0.024\pm0.011),~\rho_\pi=0.037$
for $\gamma\pp$, $x_K=0.073\pm0.015,~y_K=0.004\pm0.010,~\rho_K=0.077$
for $\gamma\kk$, and $x=0.062\pm0.011,~y=-(0.008\pm0.007),~\rho=0.058$
for the simultaneous fit.  The systematic error is taken as
the shift from $0$ plus its error. Assuming the correlation
factor is 1 between $x$ and $y$ for central value, $0.06$, $0.04$, and
$-0.63$ are obtained for $\Delta x_\pi$, $\Delta y_\pi$ and
$\rho_\pi^{sys}$, respectively; $0.08$, $0.02$, and $0.52$ are
obtained for $\Delta x_K$, $\Delta y_K$ and $\rho_K^{sys}$,
respectively; and for the simultaneous fit, $0.07$, $0.02$, and $-0.43$
are obtained for $\Delta x$, $\Delta y$ and $\rho^{sys}$,
respectively.
Studies with the MC demonstrate that a systematic error in modeling
the $\theta_{\gamma}$, $\phi_M$, or $\theta_M$ efficiency produces a
shift of $x$ and $y$ of approximately the same size in both $\chi_{c0}$
and $\chi_{c2}$ samples, where the latter sample is generated with our
nominal results for $x$ and $y$.  Therefore, we assume the observed
shift from $x = 0$ and $y = 0$ for the true $\chi_{c0}$ data is an
estimate of the systematic error on the measured values of $x$ and $y$
for radiative decays to $\chi_{c2}$.

\begin{figure}
\begin{center}
\includegraphics[height=5.5in]{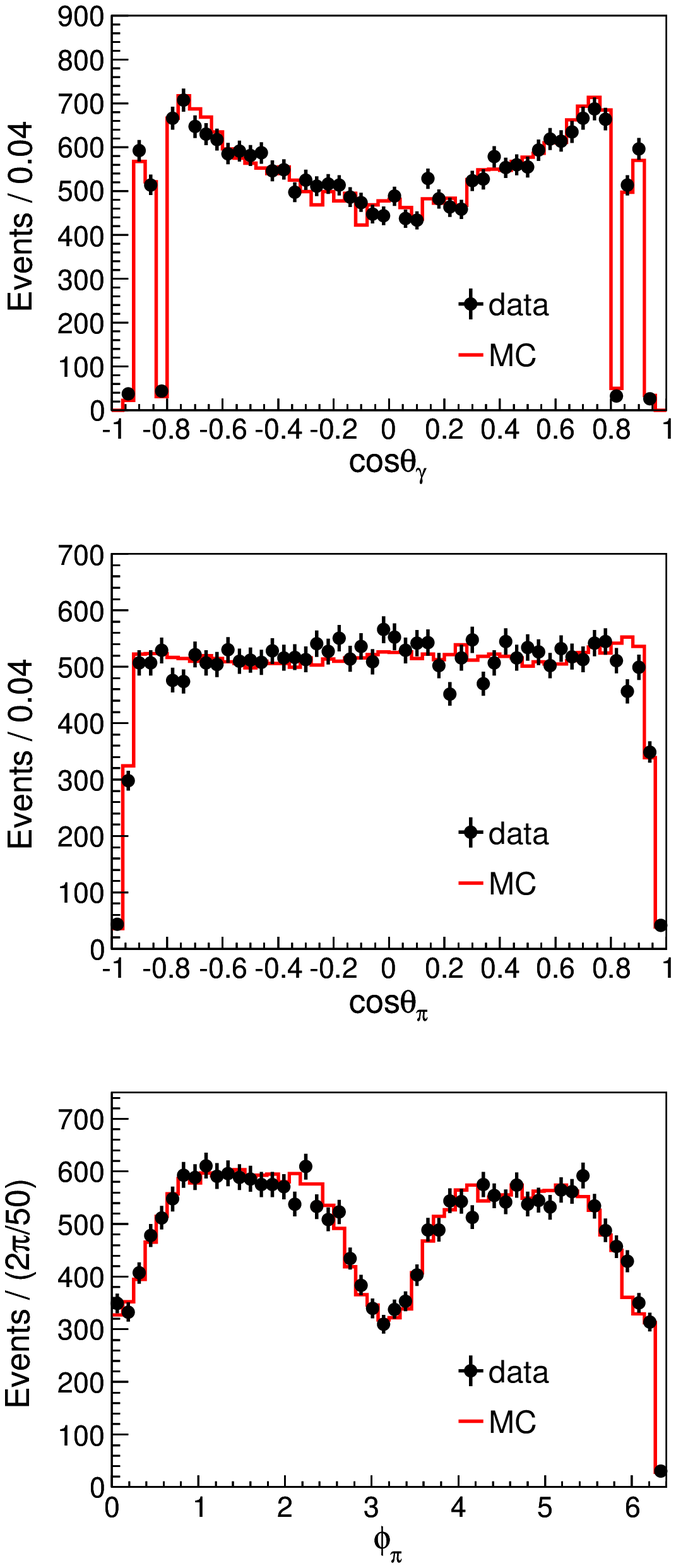}
\includegraphics[height=5.5in]{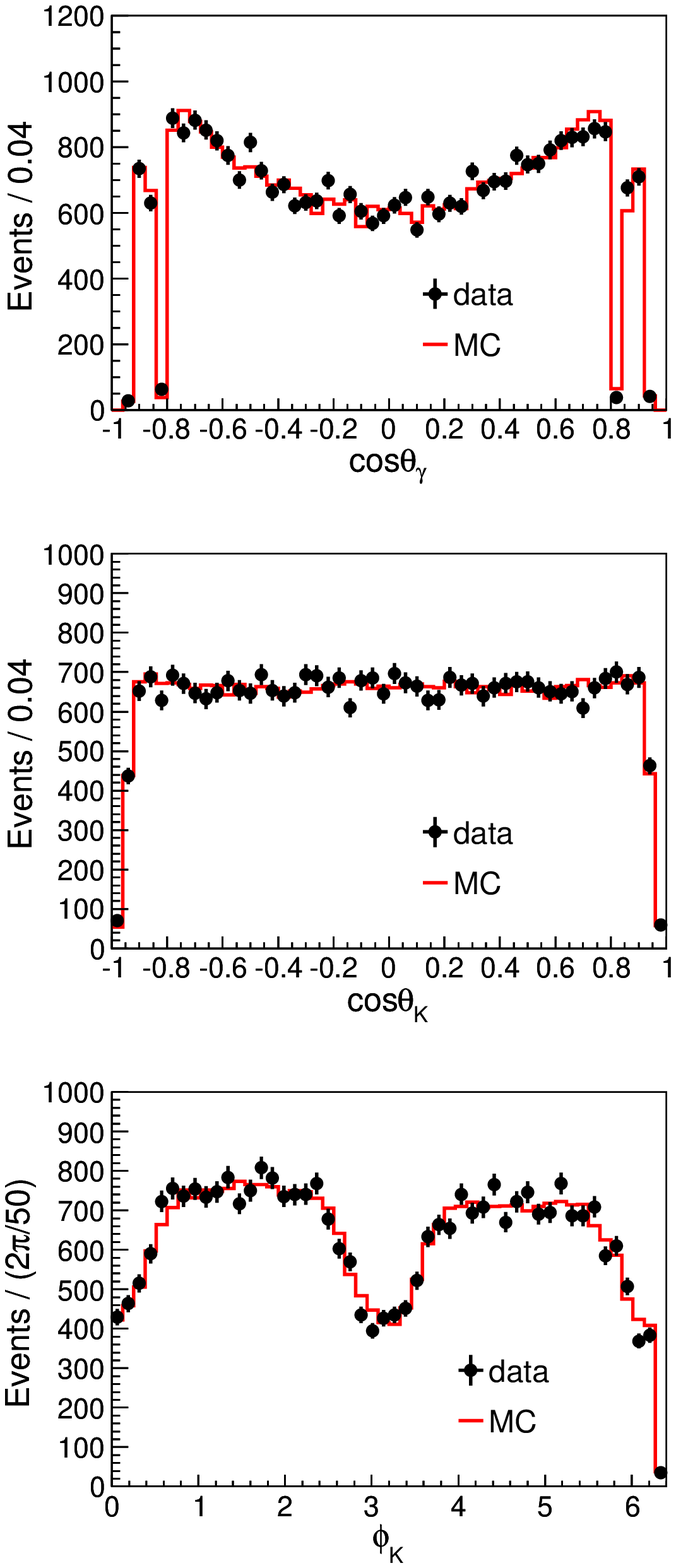}
\caption{Comparisons of $\cos\theta_\gamma$, $\cos\theta_M$, and
$\phi_M$ angular distributions between data (dots with error bars) and
MC simulation (histograms) for $\chicz\to \pp$ (left) and $\chicz\to
\kk$ (right).}\label{chic0-angle}
\end{center}
\end{figure}

The systematic error of the detector response contains
systematic errors associated with the simulation of charged track
finding, photon detection efficiency, mass resolution of $\chict$,
kinematic fit, PID efficiency, trigger efficiency, etc. Comparisons
between data and MC simulation for the angular distributions of
$\chicz$ decay events are shown in Fig.~\ref{chic0-angle}.

%%%% chi_c0 angle distribution %%%%

\subsection{\boldmath $\MM$ background in $\gamma\pp$}
In $\gamma\pp$, the dominant backgrounds are $(\gamma)\MM$ events and
continuum events, which contribute at the $1.5\%$ level.  The
$\psip\to(\gamma)\MM$ background events are estimated by MC
simulation, while the continuum events are estimated using the data
sample taken at $\sqrt{s}=3.65$ GeV.
The $\mu$/$\pi$
misidentification ratio has been checked using a control sample of
$\psip \to \pp \jpsi \to \pp \mu^+ \mu^-$ and its value is measured
to be $(1.7\pm2.6)\%$.
Considering the uncertainty from $\psip\to\MM$ branching raito (10.4\%)~\cite{pdg},
the $(\gamma)\MM$ background level is determined to be $(1.5\pm0.2)\%$.
We vary the background level
by $1\sigma$ (from $1.5\%$ to $1.7\%$ or $1.3\%$) in the fit and take
the difference of the fitted $x$ and $y$ values as the systematic
error due to the $\MM$ background uncertainty. The differences are
$\Delta x=0.02,~\Delta y=0.03$.

\subsection{\boldmath $\gamma\pp$ and $\gamma\kk$ cross contamination}
The systematic error arising from $\gamma\pp$ and $\gamma\kk$ cross
contamination is also determined by MC simulation. The $\gamma\kk$
background contamination in $\gamma\pp$ is $0.3\%$, while the
$\gamma\pp$ background contamination in $\gamma\kk$ is $0.7\%$.
Signal MC samples of $\gamma\pp$ and $\gamma\kk$ with
$x=\sqrt{3},~y=\sqrt{6}$ are generated and mixed according to the
estimated amount of cross contamination determined by MC
simulation. The differences on the fitted $x$ and $y$ values are taken
as the systematic errors due to the $\gamma\pp$ and $\gamma\kk$ cross
contamination, which are $\Delta x_\pi=0.01,~\Delta
y_\pi=0.02$ for $\gamma\pp$ and $\Delta x_K=0.01,~\Delta y_K=0.02$ for
$\gamma\kk$.

\subsection{\boldmath $\chicz$ contamination}

There are some $\chi_{c0}$ events in the $\chi_{c2}$
signal region due to the overlap of the $\pp/\kk$ invariant mass
peaks. The contamination from $\chi_{c0}\to \pp$
background events is only about $0.7\%$ for $\chi_{c2} \to \pp$, and
the contamination from $\chi_{c0}\to \kk$ background events is
$1.1\%$ for $\chi_{c2} \to \kk$ according to MC simulations.
Signal MC samples of $\chicz\to\pp/\kk$ are generated and mixed
into $\chict\to\pp/\kk$ signal MC samples
according to the estimated
contamination ratio determined by MC simulation. The differences on the fitted
$x$ and $y$ values from the input ones are taken as the systematic errors.
The systematic uncertainties due to
$\chi_{c0}$ contamination can be ignored for $\chi_{c2} \to \pp$,
while they are $\Delta x_K=0.01,~\Delta y_K=0.02$ for $\chi_{c2} \to
\kk$.

\subsection{Total systematic error}
The systematic error sources discussed above are summarized in
Table~\ref{total-err}. Here the correlation coefficients ($\rho_\pi$
and $\rho_K$) from background uncertainties including $\MM$
background, $\pi/K$ cross contamination background, and $\chi_{c0}$
background contamination are set to be 1, and the total correlation
coefficient $\rho$ is calculated as
$\rho=\sum_i\frac{\rho_i\sigma_{xi}\sigma_{yi}}{\sigma_x\sigma_y}$,
where $i$ runs over all systematic errors. The total systematic
errors are $\Delta x_\pi=0.07,~\Delta y_\pi=0.06$ in $\gamma\pp$,
$\Delta x_K=0.08,~\Delta y_K=0.04$ in $\gamma\kk$, and $\Delta
x=0.07,~\Delta y=0.05$ in the simultaneous fit.

%%%%%%%%% Table III %%%%%%%%%%%%
\begin{table}
\begin{center}
\caption{Summary of the systematic errors and correlations.} \label{total-err}
\begin{tabular}{lcccccc}
  \hline\hline
  Source & $x_\pi$ &  $y_\pi$ & $\rho_\pi$ & $x_K$ & $y_K$ & $\rho_K$\\
  \hline
  MC simulation & $0.06$ & $0.04$ & $-0.63$ & $0.08$ & $0.02$ & $0.52$\\
  $\MM$ background & $0.02$ & $0.03$ & $1$ & - & - & -\\
  $\pi/K$ cross contamination & $0.01$ & $0.02$ & $1$ & $0.01$ & $0.02$ & $1$\\
  $\chi_{c0}$ contamination & - & - & - & $0.01$ & $0.02$ & $1$\\
  Total & $0.07$ & $0.06$ & $-0.17$ & $0.08$ & $0.04$ & $0.38$\\
  \hline\hline
\end{tabular}
\end{center}
\end{table}

\section{Conclusion and Discussion}
\begin{figure}
\begin{center}
\includegraphics[height=2.5in]{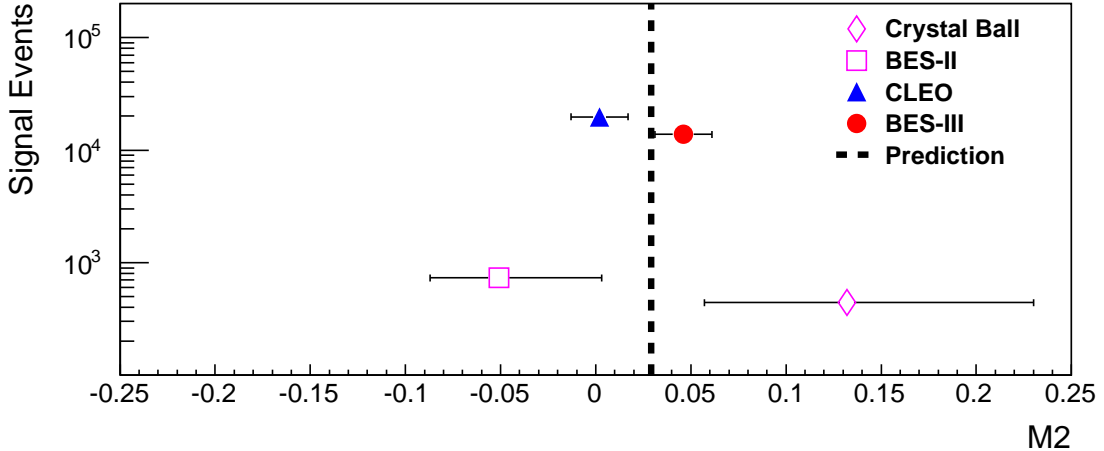}
\caption{ Experimental measurements for the normalized M2 amplitude
together with theoretical prediction~\cite{mult-pole,prediction}
assuming the charm quark mass to be $1.5$ GeV/c$^2$ and no anomalous
magnetic moment. CLEO's result is from free E3 amplitude fit.}\label{m2-signal}
\end{center}
\end{figure}

The final helicity amplitude results are:

\begin{equation}
x_\pi=1.55^{+0.08}_{-0.07}\pm0.07,~y_\pi=2.06^{+0.10}_{-0.09}\pm0.06,~\rho_\pi^{stat}=0.890,~\rho_\pi^{sys}=-0.17
\end{equation}
for $\gamma\pp$,
\begin{equation}
x_K=1.55\pm0.08\pm0.08,~y_K=2.13^{+0.11}_{-0.10}\pm0.04,~\rho_K^{stat}=0.902,~\rho_K^{sys}=0.38
\end{equation}
for $\gamma\kk$, and
\begin{equation}
x=1.55\pm0.05\pm0.07,~y=2.10\pm0.07\pm0.05,~\rho^{stat}=0.896,~\rho^{sys}=0.26
\end{equation}
from the simultaneous fit, where the first errors are statistical and
the second systematic. $\rho^{stat}$ and $\rho^{sys}$ are the
correlation coefficients between $x$ and $y$ of the statistical and
systematic errors. Then the normalized M2 and E3 amplitude are determined to be:
\begin{equation}
M2=0.046\pm0.010\pm0.013,~E3=0.015\pm0.008\pm0.018
\end{equation}
where the first errors are statistical and the second systematic.  By
investigating the difference of the logarithmic likelihoods between a
pure E1 transition and the best nominal fit, the statistical
significance of the M2 amplitude contribution is estimated to be
$4.4\sigma$, which means evidence of the M2 contribution has been
observed. As for the E3 signal, the current measurement is consistent
with zero. The M2 experimental results from different measurements are
shown in Fig.~\ref{m2-signal}. Our measurement agrees with prediction
and is consistent with CLEO's result within $2\sigma$ when E3 is
free~\cite{cleo} and BESII's result~\cite{bes2} within $1.7\sigma$.

%%%% M2 signal %%%%

%\section{Summary}
In summary, the higher order multipole amplitudes in
$\psip\to\gamma\chict\to\gamma\pp/\gamma\kk$ are studied with the
BESIII experiment based on $(1.06\pm0.04)\times10^8$ $\psip$ events.
Evidence of an M2 amplitude is observed. This measurement agrees with
prediction and is consistent with the charm quark having no anomalous
magnetic moment~\cite{mult-pole,prediction,quarkmass}.

\section{Acknowledgement}

The BESIII collaboration thanks the staff of BEPCII and the computing center for their hard efforts. This work is supported in part by the Ministry of Science and Technology of China under Contract No. 2009CB825200; National Natural Science Foundation of China (NSFC) under Contracts Nos. 10625524, 10821063, 10825524, 10835001, 10935007; the Chinese Academy of Sciences (CAS) Large-Scale Scientific Facility Program; CAS under Contracts Nos. KJCX2-YW-N29, KJCX2-YW-N45; 100 Talents Program of CAS; Istituto Nazionale di Fisica Nucleare, Italy; Siberian Branch of Russian Academy of Science, joint project No 32 with CAS; U. S. Department of Energy under Contracts Nos. DE-FG02-04ER41291, DE-FG02-91ER40682, DE-FG02-94ER40823; University of Groningen (RuG) and the Helmholtzzentrum fuer Schwerionenforschung GmbH (GSI), Darmstadt; WCU Program of National Research Foundation of Korea under Contract No. R32-2008-000-10155-0

\end{document}